\documentclass[12pt]{iopart}

\usepackage{iopams}
\usepackage{soul}										
\setstcolor{red}	  
\usepackage{cite}

\usepackage{epsfig}
\usepackage{amsmath,amssymb}
\usepackage{bm}
\usepackage{graphicx}
\usepackage{cancel}
\usepackage{empheq}
\usepackage{fancyvrb}
\usepackage{ulem}
\usepackage{booktabs}
\usepackage{hyperref}
\usepackage{lineno}

\usepackage{color}
\usepackage{multirow}
\usepackage{makecell}

\begin{document}


\title{Upper Limit of Fusion Reactivity in Laser-Driven $p+{^{11}{\rm B}}$ Reaction }

\author{Eunseok Hwang}%
\ead{hwangeunseok94@gmail.com}
\address{
        Department of Physics and OMEG Institute, 
        Soongsil University, 
        Seoul, 
        06978, 
        Republic of Korea
}

\author{Myung-Ki Cheoun}%
\ead{cheoun@ssu.ac.kr}
\address{
        Department of Physics and OMEG Institute, 
        Soongsil University, 
        Seoul, 
        06978, 
        Republic of Korea
}

\author{Dukjae~Jang \footnote{Corresponding Author}}%
\ead{dukjaejang91@gmail.com}
\address{
     Department of Physics, 
     Gachon University, 
     Gyeonggi-do, 
     13120, 
     Republic of Korea}
\address{
     School of Liberal Arts, 
     Korea University of Technology and Education, 
     Chungcheongnam-do,
     31253, 
     Republic of Korea
}
\address{
      Department of Physics, 
      Incheon National University, 
      Incheon, 
      22012, 
      Republic of Korea
}

\begin{abstract}
We explore the averaged fusion reactivity of the $p+{^{11}{\rm B}}$ reaction in tabletop laser experiments using a plasma expansion model. We investigate the energy distribution of proton beams accelerated by lasers as a function of electron temperature $T_e$, characterizing the sheath field generated by fast electrons at the rear side of a laser-irradiated thin foil (pitcher target), and the dimensionless acceleration time $\omega_{pi} t_{\rm acc}$, where $\omega_{pi}$ is the ion plasma frequency. By combining these distributions with the fusion cross-section, we identify the optimal conditions that maximize the fusion reactivity, with $\left\langle \sigma v \right\rangle = 8.12 \times 10^{-16}\,{\rm cm^3/s}$ at $k_B T_e = 10.0\,{\rm MeV}$ and $\omega_{pi} t_{\rm acc} = 0.503$. These findings suggest that an upper limit exists for the fusion reactivity achievable in laser-driven $p+{^{11}{\rm B}}$ fusion experiments.
\end{abstract}

\maketitle

\date{\today}

\maketitle

\section{Introduction \label{S1}}
Laser fusion is a promising pathway to clean energy, with deuterium-tritium (D+T) fusion long regarded as the leading candidate due to its high reaction cross-section at achievable temperatures \cite{nuckolls_laser_1972, lindl_development_1995}. However, the neutron flux generated in D+T fusion introduces significant challenges, including radiation hazards and nuclear waste. As one possible solution to address this issue, the aneutronic $p+^{11}{\rm B}$ reaction has been studied. In particular, recent advancements in high-power laser experiments \cite{STRICKLAND1985219} have enabled the acceleration of proton beams \cite{PhysRevLett.73.1801, PhysRevLett.84.4108, PhysRevLett.85.2945, 2006NatPh...2...48F}, making it possible to realize fusion using tabletop laser systems.

Despite extensive efforts to realize laser-driven $p+{^{11}\rm B}$ fusion, its practical implementation in tabletop laser setups remains challenging due to the high energy threshold, relatively low reaction cross-section compared to D+T fusion at low proton energy ($\le 60\,{\rm keV}$). Several research groups worldwide have investigated this challenge \cite{PhysRevE.103.043208, Eliezer_Martinez-Val_2020, Hora_Eliezer_2017, HORA2017177}, focusing on enhancing the high-energy tail of the proton beam to achieve resonance peaks in the fusion cross-section. Non-thermal conditions in laser-driven setups have been extensively explored as a means to generate such high-energy proton distributions, leading to experimentally observed fusion yields in recent studies \cite{PhysRevE.72.026406, labaune_fusion_2013, PhysRevX.4.031030, Margarone_2015, Baccou_Depierreux_2015, PhysRevE.101.013204, 10.3389/fphy.2020.00343, PhysRevE.103.053202, mehlhorn_path_2022, magee_first_2023}, as well as measurements of the $p+{}^{11}\mathrm{B}$ cross section near the resonance energy region \cite{Zhang_2023}. Such efforts have contributed to the generation of proton beams capable of inducing nuclear reactions. In addition, the target conditions play a crucial role in achieving high $\alpha$ yields. Recently, to mitigate the effects of power deposition from alpha particles in boron target, hydrogen-doped boron targets have been employed in experiments, resulting in enhanced $\alpha$ yields \cite{Li_Wang_2024}.

Although $p+{^{11}{\rm B}}$ fusion has been investigated in various table-top laser experiments, the mechanisms for increasing fusion yields have not yet been systematically understood. One of the key factors for understanding the fusion is the fusion reactivity, $\left\langle \sigma v \right\rangle$. However, when the proton is accelerated in laser-driven experiments, the energy distribution of the beam deviates from the Maxwell-Boltzmann distribution. Thus, the fusion reactivity can also deviate from the typical value obtained under the assumption of a Maxwell–Boltzmann distribution, which is commonly employed in nuclear astrophysical processes such as the r-, s-, and p-processes. Hence, in this study, we investigate the effect of the high-energy tail of proton beams, generated via the Target Normal Sheath Acceleration (TNSA) mechanism, on the fusion cross-section resonance. Based on the TNSA process, we investigate how key laser parameters influence the high-energy tail of the accelerated proton spectrum and determine the conditions under which the averaged fusion reactivity $\left\langle \sigma v \right\rangle$ of $p+^{11}{\rm B}$ reaction is maximized. This work provides insights into experimental design for future fusion studies as laser facilities continue to advance, advancing the way for more efficient laser-driven fusion experiments.

\section{Formalism}
The yield of alpha particles in the $^{11}{\rm B} (p, \alpha)2\alpha$ reaction is given by
\begin{eqnarray}
N_\alpha = 3 n_p n_{^{11}{\rm B}} \langle \sigma v \rangle V \tau,
\label{eq_alpha}
\end{eqnarray}
where $n_p$ and $n_{^{11}{\rm B}}$ are the number densities of protons and $^{11}{\rm B}$, respectively, $\sigma$ is the reaction cross-section, $v$ is the relative velocity between the reacting particles, $V$ is the interaction volume, and $\tau$ is the interaction time. The averaged fusion reactivity, $\langle \sigma v \rangle$, is the key quantity determining the alpha particle yield and is defined as the velocity-averaged product of the $\sigma$ and the $v$:
\begin{eqnarray}
\langle \sigma v \rangle = \int \int f_1({\bf v}_1) f_2 ({\bf v}_2) \sigma v  d{\bf v}_1  d{\bf v}_2,
\label{eq_reac}
\end{eqnarray}
where $f_i({\bf v}_i)$ represents the normalized velocity distribution of particle $i$ ($i=1,2$), and $v = |{\bf v}_2 - {\bf v}_1|$ is the relative velocity.

In nuclear astrophysical processes such as r-, s-, and p-processes, thermonuclear reaction rates are typically evaluated under the assumption that the velocity distributions of the reacting particles follow the Maxwell-Boltzmann distribution. Under this assumption, Eq.\,(\ref{eq_reac}) can be transformed into the center-of-mass frame and simplified to depend only on the relative velocity distribution. Additionally, at low energies, the reaction cross-section is often expressed in terms of the astrophysical S-factor and the penetration factor, enabling further simplification of the reaction rate. Such thermonuclear reaction rates, computed under the assumption of thermal equilibrium, are compiled in libraries such as JINA REACLIB \cite{Cyburt_2010}, which are widely adopted in nuclear astrophysics for modeling astrophysical processes. However, compared to astrophysical conditions, the evaluation of Eq.\,(\ref{eq_reac}) in laser-driven fusion experiments differs from that in astrophysical environments, as the particle velocity distributions deviate from the Maxwell-Boltzmann distribution.
\begin{figure*}
\centering
\includegraphics[width=0.75\linewidth]{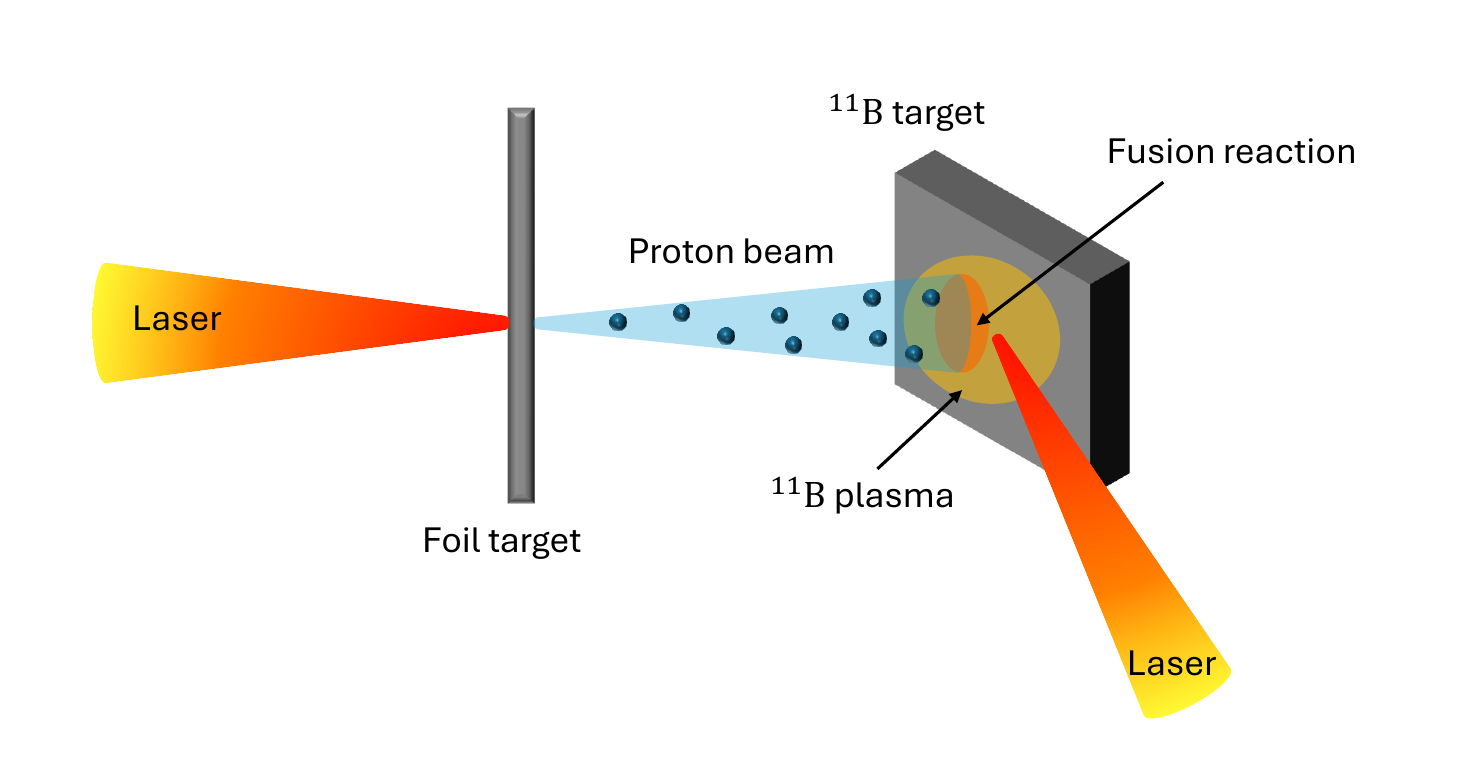}
\caption{Schematic diagram of the experimental setup for the $p+^{11}{\rm B}$ reaction in a {\it pitcher-catcher} type. Protons are accelerated from the foil target by a short-pulse laser, and the accelerated proton beam interacts with a $^{11}{\rm B}$ plasma ionized by the secondary plasma.}
\label{fig:exp_setup}

\end{figure*}

When considering the {\it pitcher-catcher} experimental type shown in Fig.~\ref{fig:exp_setup}, a short-pulse laser irradiates a thin foil target, generating a dense population of relativistic electrons. These electrons rapidly expand beyond the rear surface of the target, where the sheath surface is given by
\begin{eqnarray}
    S_{{\rm sheath}} = \pi(r_0 + d_t \tan \theta)^2,
    \label{sheath}
\end{eqnarray} 
where $d_t$ is the target thickness and $r_0$ the initial radius of the laser spot. For the angle, we adopt the typical angle of $\theta=25^\circ$, the half-angle divergence of the hot electron inside the target \cite{fuchs_laser-driven_2006}. The resulting strong sheath electric field, generated by charge separation, induces an electrostatic potential that accelerates ions along the target normal direction, a process known as the TNSA mechanism. The accelerated protons then interact with the $^{11}{\rm B}$ target, which is ionized by a secondary laser pulse. Although the fusion reactivity $\left\langle \sigma v \right\rangle$ in our study depends only on the fusion cross section and the velocity distribution of protons and is independent of the catcher target condition due to the normalization procedure explained below, the catcher target becomes important when evaluating the full reaction rate, $n_p n_{^{11}\mathrm{B}} \left\langle \sigma v \right\rangle$. In such cases, various target configurations such as cold boron, boron nitride catchers, or hydrogen-doped boron targets \cite{Li_Wang_2024} can be employed to enhance the $\alpha$ yield, in addition to fully ionized targets.

In this regime, we rewrite Eq.\,(\ref{eq_reac}) as follows:
\begin{eqnarray}
    \langle \sigma v \rangle = \int \int f_p({\bf v}_p) f_{^{11}{\rm B}} ({\bf v}_{^{11}{\rm B}}) \sigma v  d{\bf v}_p  d{\bf v}_{^{11}{\rm B}},
    \label{eq_reac_pB}
\end{eqnarray}
where ${\boldsymbol v}_p$ and ${\boldsymbol v}_{\rm ^{11}B}$ denote velocity of proton and ${\rm ^{11}B}$, respectively. The ${\rm ^{11}B}$ target is ionized by the secondary laser, which is typically a relatively low-intensity, nanosecond-duration laser ($\sim$ ns laser). In contrast, under low-energy conditions, the mass of ${\rm ^{11}B}$ is sufficiently heavy, leading to relatively much slow velocity of ${}^{11}\mathrm{B}$ compared to proton, i.e., $v_{\rm ^{11}B} \ll v_p$. Therefore, the relative velocity simplifies to $v \simeq v_p$. Furthermore, the plasma expansion model adopted in this study, as explained below, is a one-dimensional (1D) model, which simplifies the three-dimensional integration to $d{\bf v} = dv_x$ along the propagation direction. Consequently, using $f(v_p) dv_p = f(E_p) dE_p$, one can rewrite Eq.~(\ref{eq_reac_pB}) as follows:
\begin{eqnarray}
    \langle \sigma v \rangle = \int f_p(v_p) \sigma(E_r) v_p dv_p =  \sqrt{\frac{2}{m_p}} \int f_p(E_p) \sigma(E_r) \sqrt{E_p} dE_p,
\label{reac_e}    
\end{eqnarray}
where $m_p$ and $E_p$ denote the mass and energy of the proton, respectively, and $E_r = \mu v^2/2$ is the center-of-mass kinetic energy associated with the reduced mass $\mu$. Eq.\,(\ref{reac_e}) depends solely on the fusion cross-section and the distribution function of accelerated proton beams. Therefore, the ability of the proton distribution function to activate the resonance peak of the cross-section is a key factor in determining the fusion yield.

To determine $f_p({{\bf v}_p)}$, we adopt the 1D plasma expansion model proposed in Refs. \cite{Crow_Auer_Allen_1975, PhysRevLett.90.185002}. In this model, the electron distribution is assumed to follow a Maxwell-Boltzmann distribution, yielding the electron density $n_e$ as
\begin{eqnarray}
n_e = n_{e0} \exp \left( \frac{e \Phi}{k_B T_e} \right),
\label{ne}
\end{eqnarray}
where $n_{e0}$ is the unperturbed electron density, i.e., the background electron density in the absence of electrostatic potential, $\Phi$ is the electrostatic potential, and $T_e$ is the electron temperature. For the electron temperature, we employ an empirical formula based on the ponderomotive scaling \cite{PhysRevLett.69.1383}:
\begin{eqnarray}
T_e = m_e c^2 \left[ \sqrt{1 + \frac{I \lambda_\mu^2}{1.37 \times 10^{18}}} - 1 \right],
\label{Te}
\end{eqnarray}
where $m_e$ is the electron mass, $c$ is the speed of light, $I$ is the laser intensity in units of ${\rm W\,cm}^{-2}$, and $\lambda_\mu$ is the laser wavelength in micrometers. The electron density is given by $n_{e0} = N_{e}/(c \tau_{\rm laser} S_{\rm sheath})$, where the total number of electrons is defined as $N_e = f E_{\rm laser}/T_e$, with $E_{\rm laser}$ being the laser energy and $\tau_{\rm laser}$ is the duration pulse of laser. For the energy efficiency from the laser to fast electrons, $f$, we adopt $f = 1.2 \times 10^{-15} I^{0.74}\,{\rm W cm^{-2}}$ \cite{10.1063/1.872867, PhysRevE.56.4608}.

Given the electron conditions, the electrostatic potential $\Phi$, which governs ion acceleration, is obtained by solving the following Poisson equation:
\begin{eqnarray}
    \epsilon_0 \frac{\partial^2 \Phi}{\partial x^2} = e (n_e - Z n_i),
    \label{Poisson}
\end{eqnarray}
where $\epsilon_0$ is the dielectric permittivity, $Z$ is the charge number of ion, and $n_i$ is the ion number density. With the potential $\Phi$, the evolution of the ion density expanding from the foil into vacuum is described by the continuity equation and the ion equation of motion:
\begin{eqnarray}
    \left( \frac{\partial}{\partial t} + v_i \frac{\partial }{\partial x} \right) n_i &=& - n_i \frac{\partial v_i}{\partial x}, 
    \label{eq:con} \\[12pt]
    \left( \frac{\partial}{\partial t} + v_i \frac{\partial}{\partial x} \right) v_i &=& - \frac{Ze}{m_i} \frac{\partial \Phi}{\partial x},
    \label{eq:eom}
\end{eqnarray}
where $v_i$ is the ion velocity. As an initial condition, ions occupy the half-space at $t = 0$, while electrons follow a Maxwell-Boltzmann distribution in Eq.\,(\ref{ne}). Consequently, by solving coupled Eqs.~(\ref{Poisson}), (\ref{eq:con}), and (\ref{eq:eom}), the ion energy distribution can be obtained.

Assuming quasi-neutrality during plasma expansion, an analytic expression for the number of protons per unit energy can be derived, referred to as the ``self-similar solution" \cite{PhysRevLett.90.185002}. In the TNSA scheme, this expression takes the following form \cite{fuchs_laser-driven_2006}:
\begin{eqnarray}
 \frac{dN}{dE} = \frac{n_{e0} c_s t_{\rm acc} S_{\rm sheath}}{\sqrt{2 E k_B T_e}} \exp \left[  - \sqrt{\frac{2E}{k_B T_e}} \right],
 \label{N_ss}
\end{eqnarray}
where $c_s = \sqrt{Z k_B T_e / m_i}$ is the ion acoustic velocity, which can also be expressed as $c_s = \lambda_{D0}\omega_{pi}$ with the initial Debye length defined as $\lambda_{D0}=(\epsilon_0 k_B T_e/n_{e0}e^2)^{1/2}$ and the ion plasma frequency as $\omega_{pi}=(n_{e0} Z e^2/(m_i\epsilon_{0}))^{1/2}$, and $t_{\rm acc}$ represents the effective acceleration time. For $t_{\rm acc}$, we adopt $t_{\rm acc} = 1.3 \times \tau_{\rm laser}$. We note that the relation $t_{\rm acc} = 1.3 \times \tau_{\rm laser}$, while commonly employed, is not universally valid, particularly for very short-pulse lasers. To assess the sensitivity to the factor 1.3, we introduce $a_0$ as a free parameter in the relation $t_{\rm acc} = a_0 \tau_{\rm laser}$, and examine its effect under the ELFIE laser conditions at LULI. We have confirmed that the calculated fusion reactivity $\langle \sigma v \rangle$ is not sensitive to variations in $a_0$ for $a_0 \gtrsim 0.65$. Therefore, in this study, we fix $a_0 = 1.3$ for all results. However, despite the weak sensitivity to variations in $a_0$, we note that this approach remains phenomenological and is not derived from first principles. More rigorous theoretical frameworks, particularly quasi-static models based on the self-consistent solution of the Poisson equation, are presented in Refs. \cite{PEREGO201189, perego_target_2012, Passoni_2013}.

From Eq.\,(\ref{N_ss}), the normalized energy distribution of laser-accelerated protons can be obtained as follows:
\begin{eqnarray}
    f_{p,ss}(E) = \frac{1}{\sqrt{2E k_B T_e}} \exp \left[ - \sqrt{\frac{2E}{k_B T_e}} \right].
\end{eqnarray}

A limitation of the self-similar solution is that it is valid only when the initial Debye length, $\lambda_{D0}$, is much smaller than the self-similar density scale length, $c_s t$, i.e., for $\omega_{pi} t_{\rm acc} \gg 1$. This condition implies that the plasma expands slowly enough for quasi-neutrality to be maintained throughout the acceleration process. Otherwise, the solution becomes invalid, requiring a numerical determination of the ion number distribution. In particular, ultrashort-pulse and high-intensity lasers are available in recent laser facilities, where the self-similar solution cannot be applied under these conditions. Therefore, to investigate the small $\omega_{pi} t_{\rm acc}$ regime, a numerical solution for the distribution function of ions is necessary.

One important consideration when solving Eqs.~(\ref{eq:con}) and (\ref{eq:eom}) numerically is that the distribution function must be normalized, i.e., 
\begin{eqnarray}
    \int_{E_{\rm min}}^{E_{\rm max}} f_{p}(E) dE = 1,
    \label{normal}
\end{eqnarray}
where $E_{\rm max}$ and $E_{\rm min}$ are the maximum and minimum energies, respectively. The maximum energy corresponds to the front velocity, $v_{f}$, such that $E_{\rm max} = m_p v_{f}^2/2$. Conversely, the minimum energy is determined by the slowest velocity, $v_{\rm min}$, which corresponds to the velocity at the most backward position. We obtain $v_{\rm min}$ using a root-finding method to satisfy the normalization condition of the distribution function. This velocity then defines the minimum energy of the proton beam as $E_{\rm min} = m_p v_{\rm min}^2/2$. Thus, due to this normalization condition, $E_{\rm min}$ is not always zero, especially for high-intensity lasers. This is consistent with the physical interpretation that highly accelerated proton beams do not contain slow protons.

\begin{figure}
\centering
\includegraphics[width=0.75\linewidth]{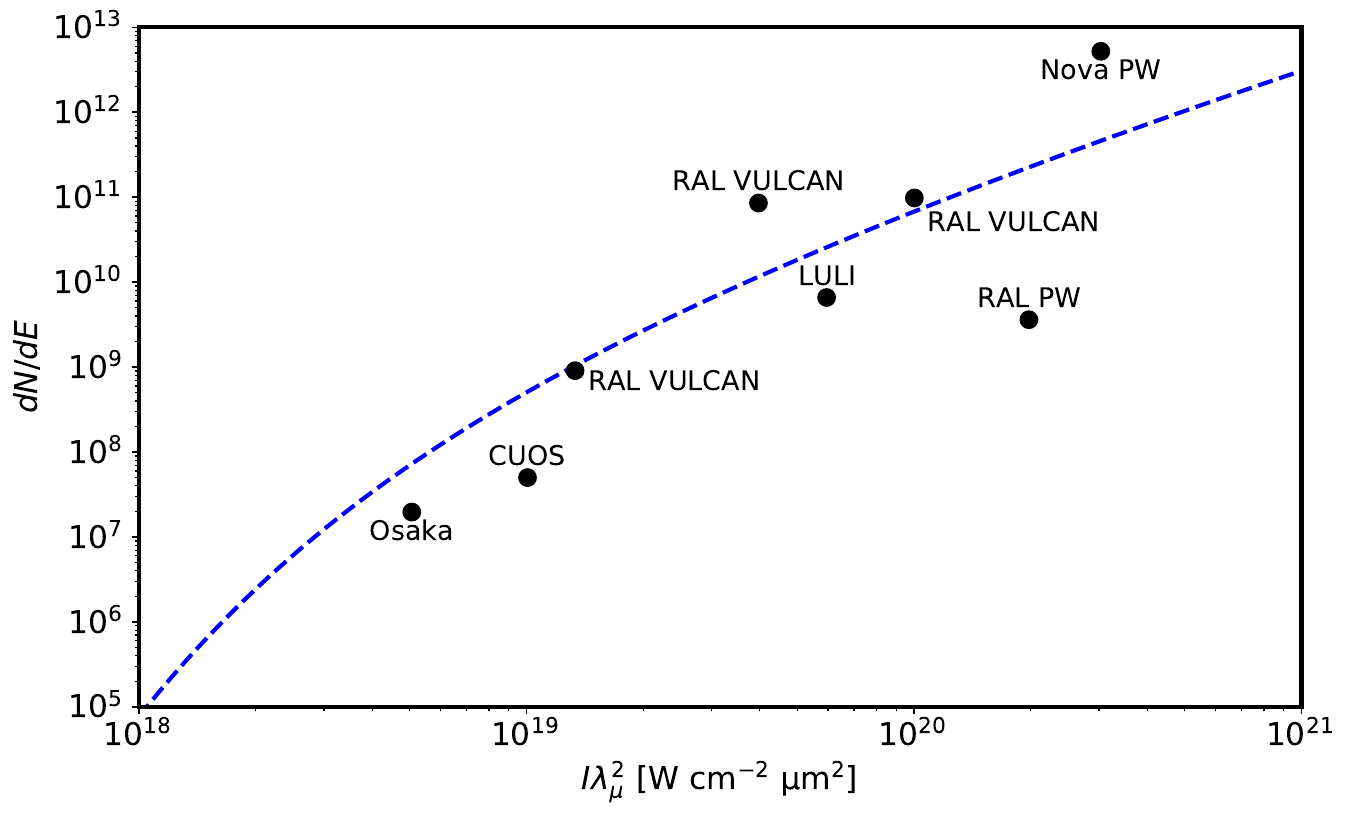}
\caption{ Proton number as a function of $I\lambda^2$, compared with published experimental data. The prediction is obtained using $\tau = 500\,{\rm fs}$, $r_0 = 10\,{\rm \mu m}$, and $d_t = 10\,{\rm \mu m}$, and by integrating the self‑similar $dN/dE$ spectrum over a 1 MeV bin centered at 10 MeV. The circular data points are extracted from Refs: Osaka\cite{osaka}, CUOS\cite{CUOS}, RAL VULCAN\cite{Ral_Vulcan1,Ral_Vulcan2}, LULI \cite{fuchs_laser-driven_2006}, RAL PW \cite{RAL_PW}, and Nova PW \cite{PhysRevLett.85.2945}. }
\label{fig_exp}
\end{figure}

Before evaluating the fusion reactivity using the proton distribution, we first assess the validity of our numerical results by comparing the calculated energy spectrum $dN/dE$ with available experimental data, as shown in Fig.\,\ref{fig_exp}. A direct comparison is not straightforward, as each experiment uses different laser parameters and integrates the proton spectrum over different energy ranges. To facilitate comparison, we follow the procedure described in Ref.\,\cite{fuchs_laser-driven_2006}, extracting the proton number by integrating the numerically obtained $dN/dE$ over a 1\,MeV bin centered at 10\,MeV. Although the absolute proton yields differ due to variations in pulse duration and integration range, the trend of $dN/dE$ as a function of $I\lambda^2$ is consistent with the experimental data and with the results reported in Ref.\,\cite{fuchs_laser-driven_2006}.

\section{Results}
\begin{figure}[t]
    \centering
    \includegraphics[width=0.75\linewidth]{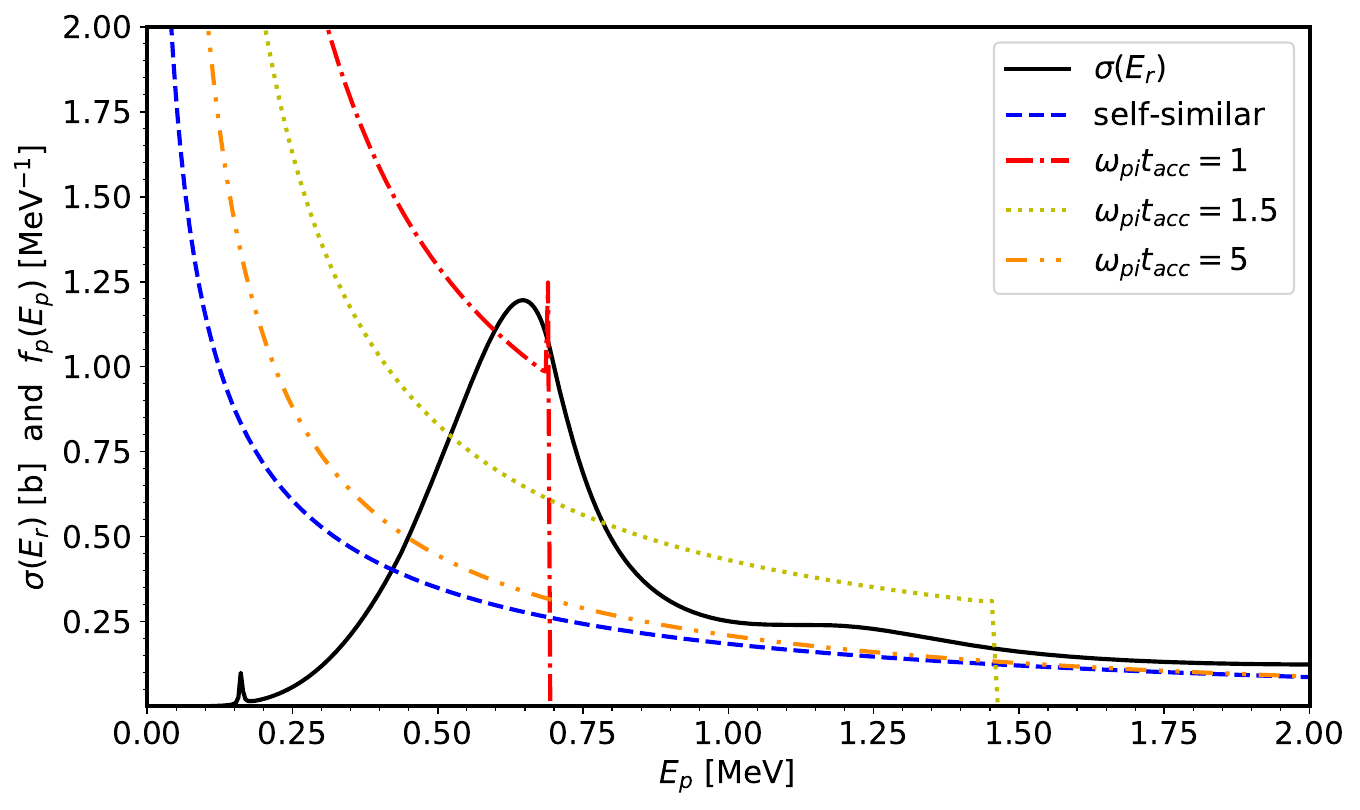}
    \caption{Fusion cross-section for the $p+^{11}{\rm B}$ reaction \cite{Nevins_2000} and distribution function as a function of $E_p$. The black solid line represents $\sigma(E_r)$ with $E_r \simeq (\mu/m_p) E_p$, while the blue dashed line corresponds to $f_p(E_p)$ obtained from the self-similar solution. For the $f_p(E_p)$ numerically obtained, the red dash-dotted, yellow dotted, and orange dash-dot-dot lines indicate $\omega_{pi} t_{\rm acc} = 1$, $1.5$, and $5$, respectively. For all $f(E_p)$ calculations, $k_B T_e = 2\,{\rm MeV}$ is adopted. }
    \label{fig:dndE}
\end{figure}

Figure \ref{fig:dndE} presents the fusion cross-section and the energy distribution of protons, which corresponds to the integrand in Eq.~(\ref{reac_e}), for $k_B T_e = 2\,{\rm MeV}$. As explained above, a large $\omega_{pi} t_{\rm acc}$ indicates that the plasma expands while maintaining quasi-neutrality. Consequently, for $\omega_{pi} t_{\rm acc} \gg 1$, $f_p(E_p)$ can be approximated by the self-similar solution.  However, for $\omega_{pi} t_{\rm acc} \lesssim 1$, this approximation breaks down, which requires numerical calculations. Furthermore, as $\omega_{pi} t_{\rm acc}$ increases, the high-energy tail of $f_p(E_p)$ is enhanced, thereby activating the resonance energy of the fusion cross-section.

Another notable feature in Fig.~\ref{fig:dndE} is that as $\omega_{pi} t_{\rm acc}$ decreases, the high-energy tail becomes shorter, but its magnitude increases. As a result, when the enhanced high-energy tail overlaps with the resonance energy region, the fusion reactivity increases. However, if $\omega_{pi} t_{\rm acc}$ is too small, the high-energy tail is truncated before reaching the resonance energy region, leading to a decrease in reaction rate. Consequently, for a fixed electron temperature $T_e$, that is, for fixed laser intensity $I$ and wavelength $\lambda_\mu$, where there exists an optimal value of $\omega_{pi} t_{\rm acc}$ that maximizes the fusion rate. From the perspective of laser parameters, this implies that there exists an optimal pulse duration $\tau_{\rm laser}$ that maximizes the fusion rate for given $I$ and $\lambda_\mu$.

\begin{figure}[t]
    \centering
    \includegraphics[width=0.75\linewidth]{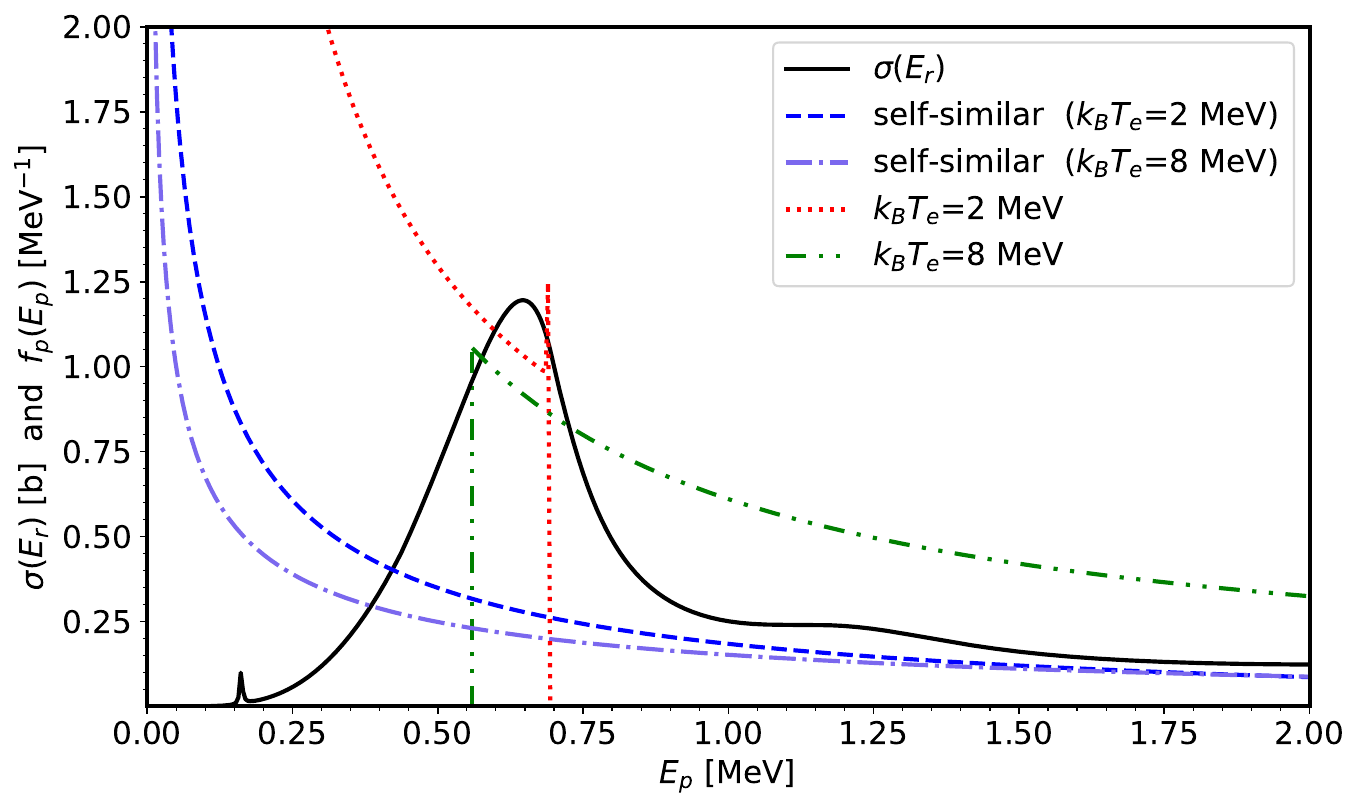}
    \caption{This figure is the same as Fig.~\ref{fig:dndE_wt=1}, but here, $f_p(E_p)$ is presented for different temperatures while keeping $\omega_{pi} t_{\rm acc} = 1$ fixed. The black solid line represents $\sigma(E_r)$ with $E_r \simeq (\mu/m_p) E_p$, while the blue dashed and blue dash-dotted lines correspond to $f_p(E_p)$ obtained from the self-similar solution for $k_B T_e = 2\,{\rm MeV}$ and $k_B T_e = 8\,{\rm MeV}$, respectively. For the numerically obtained $f_p(E_p)$, the red dotted and green dash-dot-dot lines indicate $k_B T_e = 2\,{\rm MeV}$ and $8\,{\rm MeV}$, respectively.}
    \label{fig:dndE_wt=1}
\end{figure}

Figure \ref{fig:dndE_wt=1} depicts $f_p(E_p)$ and $\sigma(E_r)$ for the same $\omega_{pi} t_{\rm acc}$ but different $T_e$. As shown in this figure, an increase in $T_e$ shifts $f_p(E_p)$ toward the high-energy region. However, due to the normalization condition in Eq.~(\ref{normal}), $f_p(E_p)$ lacks a low-energy region, implying that there are no low-velocity protons in the accelerated beams. As a result, the proton distribution function bypasses the resonance energy if $T_e$ is too high, thereby reducing fusion yields. Consequently, at very high laser intensities, where the resulting electron temperature at the foil target becomes large, the fusion yield may decrease. This implies that, for a given laser pulse duration $\tau_{\rm laser}$, there exists an optimal laser intensity $I$ that maximizes the fusion rate.

\begin{figure}[t]
    \centering
    \includegraphics[width=0.75\linewidth]{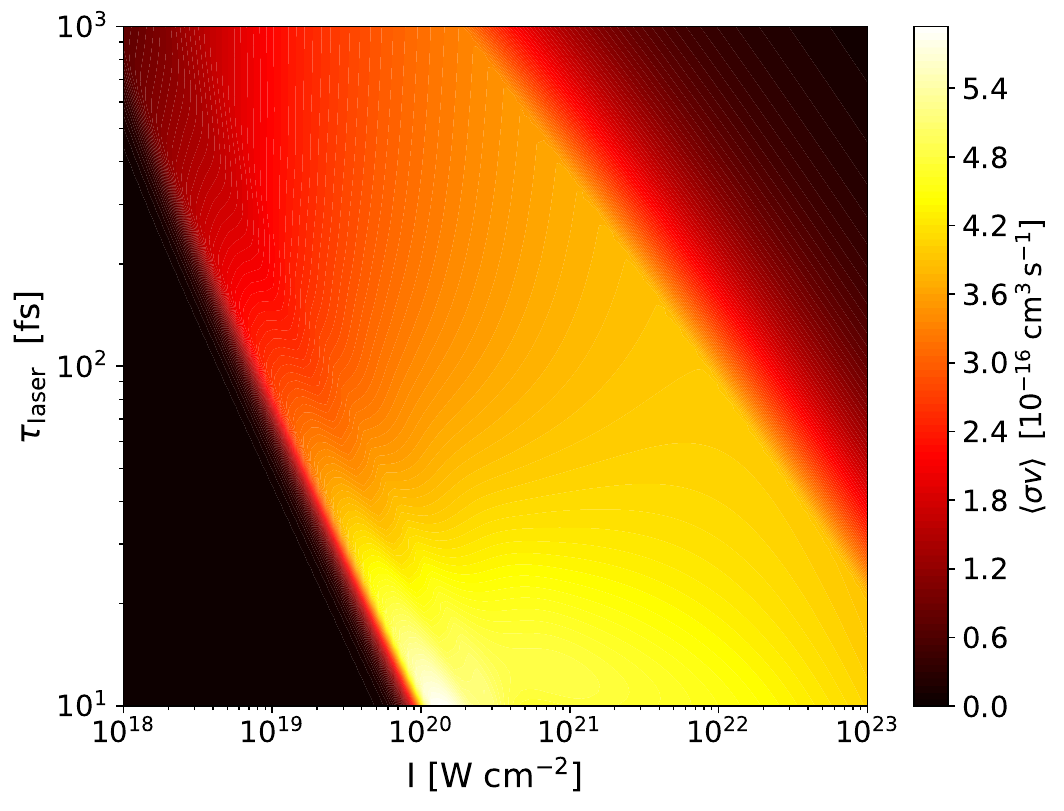}
    \caption{Averaged fusion reactivity $\left\langle \sigma v \right\rangle$ of $p+{^{11}{\rm B}}$ reaction as a function of $\tau_{\rm laser}$ and $I$. For this calculation, we adopt $\lambda = 1\,{\rm \mu m}$, $r_0 = 10\,{\rm \mu m}$, and $d_t = 10\,{\rm \mu m}$.
    }
    \label{fig:I_tau}
\end{figure}

For the parameter space of $10^{18}\,{\rm W \, cm^{-2}} \le I \le 10^{23}\,{\rm W\,cm^{-2}}$ and $10\,{\rm fs} \le \tau_{\rm laser} \le 10^3\,{\rm fs}$, we investigate the trend of $\left\langle \sigma v \right\rangle$, which is presented in Fig.~\ref{fig:I_tau}. In Fig.~\ref{fig:I_tau}, the laser wavelength $\lambda$ is fixed at $1\,\mu {\rm m}$. Although $\lambda$ varies slightly in practical applications, we find that its sensitivity is relatively small compared to $I$ and $\tau_{\rm laser}$. Moreover, in our calculation, we fix the laser spot size at $r_0 = 10\,\mu{\rm m}$ and the target thickness at $d_t = 10\,\mu{\rm m}$. Taking all these conditions into account, we obtain $\left\langle \sigma v \right\rangle$ as a function of $\tau_{\rm laser}$ and $I$, as shown in Fig.~\ref{fig:I_tau}.
In Fig.~\ref{fig:I_tau}, $\left\langle \sigma v \right\rangle$ exhibits a peak at specific values of $I$ and $\tau_{\rm laser}$.

As explained above, this behavior arises from two competing effects. First, when $\omega_{pi} t_{\rm acc}$ is too short, the resulting proton spectrum develops a strong high-energy tail but suffers from a cutoff at low energies, reducing the overlap with the fusion cross-section. Second, excessively high laser intensity $I$ leads to a high electron temperature $T_e$, which increases the minimum proton energy $E_{\rm min}$ beyond the resonance peak of the cross-section, thereby suppressing the reaction rate. Consequently, for $\lambda=1\,{\rm \mu m}$, we find that the maximum value of $\left\langle \sigma v \right\rangle$ is $5.92 \times 10^{-16}\,{\rm cm^3 s^{-1}}$ at $I = 1.32 \times 10^{20} \,{\rm W \, cm^{-2}}$ and $\tau_{\rm laser} = 10\,{\rm fs}$.

\begin{figure}[t]
    \centering
    \includegraphics[width=0.75\linewidth]{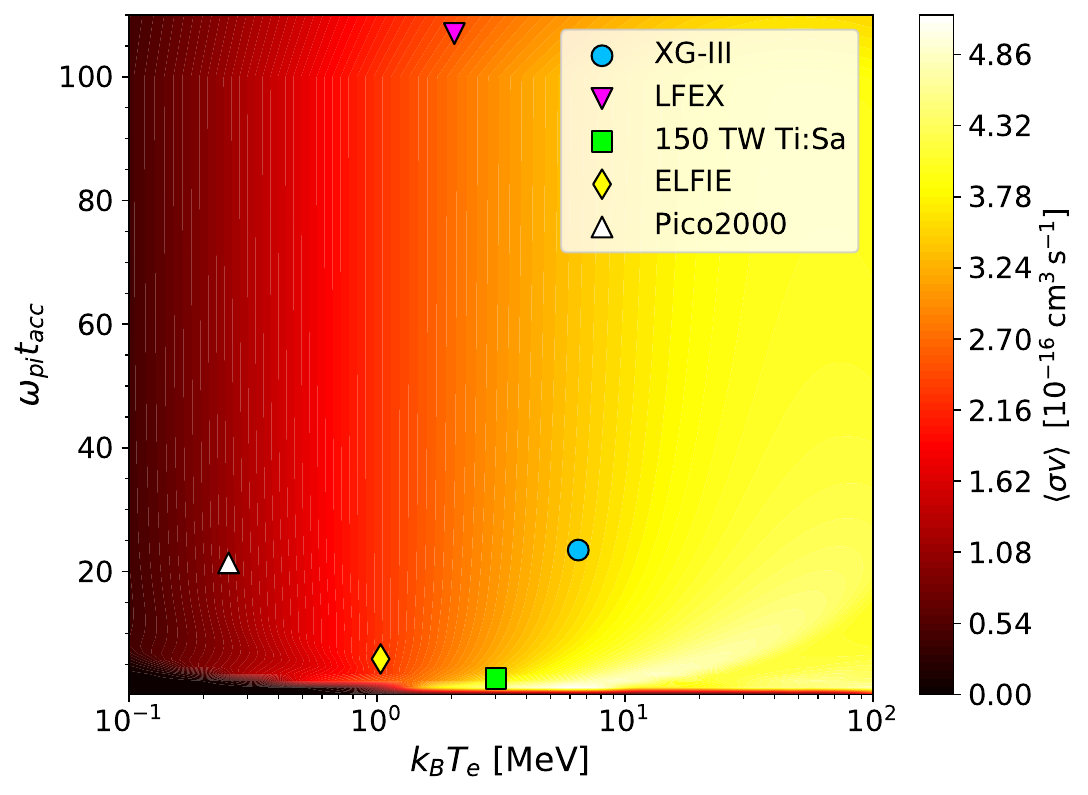}
    \caption{ The nuclear reaction rate $\left\langle \sigma v \right\rangle$ as a function of $\omega_{pi} t_{\rm acc}$ and $k_B T_e$. Each symbol represents the $\omega_{pi} t_{\rm acc}$ and $k_B T_e$ values obtained from the experimental setup: the blue circle, purple inverted triangle, green square, yellow diamond, and white triangle correspond to laser parameters for XG-III laser in Laser Fusion Research Center (LFRC) \cite{wei2023(XG-III)}, LFEX laser in Institute of Laser Engineering (ILE) \cite{10.3389/fphy.2020.00343}, 150 TW Ti:Sa laser in Raja Ramanna Centre for Advanced Technology (RRCAT) \cite{Tayyab_2019}, ELFIE laser in LULI \cite{Baccou_Depierreux_2015}, and Pico2000 laser in LULI \cite{labaune_fusion_2013}, respectively. } 
    \label{fig:wt_Te}
\end{figure}

Although $I$, $\tau_{\rm laser}$, $\lambda$, $r_0$, and $d_t$ are involved in determining the averaged fusion reactivity, as shown in Eqs.\,(\ref{Te}) and (\ref{N_ss}), they can be reduced to two parameters: $k_B T_e$ and $\omega_{pi} t_{\rm acc}$. In Fig.~\ref{fig:wt_Te}, we present $\left\langle \sigma v \right\rangle$ as a function of these two parameters. Within the given parameter space, the maximum fusion reactivity is found to be $\left\langle \sigma v \right\rangle = 8.12 \times 10^{-16}\,{\rm cm^3/s}$ at $k_B T_e = 10.0\,{\rm MeV}$ and $\omega_{pi} t_{\rm acc} = 0.503$. These values represent the optimum conditions for maximizing the fusion reactivity, while also suggesting an upper limit to the achievable reactivity for the $p+^{11}{\rm B}$ reaction in a {\it pitcher-catcher} type experimental configuration. 

In Fig.~\ref{fig:wt_Te}, we also mark the parameters estimated from previous experiments on $p+^{11}{\rm B}$ fusion in the {\it pitcher–catcher} type. Among the five parameter sets used in previous experiments, the parameters employed in Ref.\cite{Tayyab_2019} yield the highest $\left\langle \sigma v \right\rangle$. However, this does not necessarily imply that these parameters result in the highest alpha yields. As shown in Eq.(\ref{eq_alpha}), the alpha yield depends not only on the averaged fusion reactivity but also on the number of $^{11}{\rm B}$ target nuclei, the interaction time, and the interaction volume. Consequently, if the target condition can be experimentally controlled, the result of $\left\langle \sigma v \right\rangle$ shown in Fig.~\ref{fig:wt_Te} can serve as a practical guide for optimizing the fusion yield.

The electron screening effect also plays an important role in nuclear fusion. If the ${\rm ^{11}B}$ target is fully or partially ionized by electrons, electron shielding in the plasma reduces the Coulomb barrier, thereby enhancing the fusion reaction rates. To describe the electron screening effect, the weak screening approximation is widely accepted. However, the plasma produced by the laser exists in a state of high density and low temperature, meaning that the weak screening condition may not be applicable. For example, in Ref.\,\cite{wei2023(XG-III)}, which investigates ${\rm ^{11}B} + p$ fusion driven by a high-power laser, the target conditions are given as $n_{e}=4\times10^{20}\,\rm{cm^{-3}}$ and $k_BT=17\, \rm{eV}$. Under these conditions, the plasma coupling parameter is given by  $\Gamma=n_e^{1/3}(Ze)^2/(k_B T)=1.56$, which indicates that the plasma is strongly coupled. In such a non-ideal plasma, the weak screening condition is not valid. Nevertheless, since screening effects can enhance the fusion probability, their role in this regime should be further investigated in future studies.

\section{Conclusion}
In this study, we investigate the averaged fusion reactivity $\left\langle \sigma v \right\rangle$ of the reaction $p+^{11}{\rm B}$ in a laser-driven experiment of {\it pitcher-catcher} type. We present that there exists an optimal set of laser parameters, specifically the intensity $I$ and the pulse duration $\tau_{\rm laser}$ that maximizes the averaged fusion reactivity. For a fixed wavelength of $\lambda = 1\,\mu{\rm m}$, the optimal condition was found to be $I = 1.32 \times 10^{20}\,{\rm W/cm^2}$ and $\tau_{\rm laser} = 10\,{\rm fs}$.

More generally, the laser and target parameters can be reduced to two parameters of $\omega_{pi} t_{\rm acc}$ and $T_e$. For these parameters, $\omega_{pi} t_{\rm acc}$ governs the low energy cutoff, while $T_e$ controls the high energy extent of the proton energy distribution. Thus, the optimal condition emerges when these two competing effects shape the energy tail to coincide with the resonance region of the fusion cross-section. Under this condition, we find that the maximum fusion reactivity, $\left\langle \sigma v \right\rangle = 8.12 \times 10^{-16}\,{\rm cm^3/s}$, is obtained at $k_B T_e = 10.0\,{\rm MeV}$ and $\omega_{pi} t_{\rm acc} = 0.503$. A comparison with previously reported experimental parameter sets is summarized in Table~\ref{tab:1}. These results present the optimal parameters for maximizing the fusion reactivity under a laser-driven proton beam distribution. At the same time, they indicate that there exists an upper limit to the fusion reactivity, even with increased laser intensity and reduced pulse duration in $p+^{11}{\rm B}$ reactions.
\begin{table}[htbp]
\centering
\caption{Experimental laser parameters and evaluated quantities for $p + {}^{11}\mathrm{B}$ fusion. 
The second to sixth rows list the laser parameters $E_{\rm laser}$, $I$, $\tau_{\rm laser}$, $\lambda_\mu$, and $d_t$ employed in each referenced experiment, from which $k_B T_e$, $\omega_{pi} t_{\rm acc}$, and $\langle \sigma v \rangle$ are evaluated. 
The parameter $r_0$ is obtained from $E_{\rm laser}$, $I$, and $\tau_{\rm laser}$ assuming a Gaussian profile, with $r_0$ taken as the full width at half maximum (FWHM) of the laser spot size.
The last row presents the theoretical optimal values of $k_B T_e$ and $\omega_{pi} t_{\rm acc}$ that yield the maximum fusion reactivity.}
\label{tab:1}
\resizebox{\textwidth}{!}{%
\begin{tabular}{llccccc|ccc}
\toprule
\textbf{Where} & \textbf{Laser} 
& \multicolumn{5}{c|}{\textbf{Experimental Parameters}} 
& \multicolumn{3}{c}{\textbf{Evaluated Quantities}} \\
\cmidrule(lr){3-7} \cmidrule(lr){8-10}
& & $E_{\rm laser}$ [J] & $I$ [$\mathrm{W/cm^2}$] & $\tau_{\rm laser}$ [fs] & $\lambda_\mu$ [$\mu$m] 
&  $d_t$ [$\mu$m] 
& $k_B T_e$ [MeV] & $\omega_{pi} t_{\rm acc}$ 
& $\left\langle \sigma v \right\rangle$ [$10^{-16}$ cm$^3$\,s$^{-1}$] \\
\midrule
LFRC \cite{wei2023(XG-III)}        & XG-III        & 120   & $2.3\times10^{20}$ & 800   & 1.053  & 10  & 6.48  & 23.5  & 3.48 \\
ILE \cite{10.3389/fphy.2020.00343} & LFEX          & 1400  & $3.0\times10^{19}$ & 2700  & 1.05   & 25  & 2.05  & 107   & 2.56 \\
RRCAT \cite{Tayyab_2019}           & 150 TW Ti:Sa  & 2.5   & $1.0\times10^{20}$ & 25    & 0.8    & 1   & 3.02  & 2.74  & 3.58 \\
LULI \cite{Baccou_Depierreux_2015} & ELFIE         & 12    & $1.0\times10^{19}$ & 350   & 1.056  & 20  & 1.03  & 5.84  & 2.28 \\
LULI \cite{labaune_fusion_2013}    & Pico2000      & 20    & $6.0\times10^{18}$ & 1000  & 0.53   & 20  & 0.25  & 21.4  & 1.09 \\
\midrule
\multicolumn{7}{c|}{\makecell[c]{Theoretical Optimum \\ (This work)}} & 10.0 & 0.503 & 8.12 \\
\bottomrule
\end{tabular}%
}
\end{table}

We note that this work represents a foundational analysis, which can be extended by more rigorous theoretical and numerical investigations. First, our calculation is based on a fluid expansion model. Although this model is widely adopted due to its simplicity, it has limitations in cases involving ultra-short laser pulses or thicker targets. For this reason, the comparison presented in Table~\ref{tab:1} might leave room for refinement in future work. For example, to better capture such effects on $\langle \sigma v \rangle$, more advanced ion acceleration models within the TNSA framework, such as quasi-static models \cite{PhysRevLett.101.115001}, can be employed in future studies. Second,  we do not consider the dynamics of the proton distribution after its generation. Once the proton beam is produced from a thin foil target, its energy distribution may evolve as it interacts with the boron target. In particular, if the boron target is sufficiently thick, the accelerated protons can lose energy through collisions, leading to a modification of the proton energy spectrum. Such changes may in turn affect the value of $\left\langle \sigma v \right\rangle$. A detailed treatment of the beam dynamics, including energy loss and scattering processes, would require fully kinetic simulations such as Particle-in-Cell (PIC) or Monte Carlo simulations, which are beyond the scope of the present work and will be addressed in future studies.

We also discuss the impact of electron screening effects on non-equilibrium fusion reactions. Although the Salpeter enhancement factor suggests a significant increase in the reaction rate due to electron screening, its applicability under laser-driven, non-equilibrium conditions remains uncertain. Future studies should focus on refining theoretical models to account for these effects more accurately and exploring experimental validation of the predicted optimal conditions. Finally, such a theoretical framework, along with our systematic investigation of laser parameters, would contribute not only to advancing alternative clean energy sources but also to nuclear reaction rate measurements on tabletop laser systems for nuclear astrophysics.

\ack
The authors appreciate the anonymous referees for their valuable suggestions, which contributed to improving the manuscript.
The work of EH and MKC was supported by the Basic Science Research Program of the National Research Foundation of Korea (NRF) under Grants No.\,RS-2021-NR060129.

\vspace{1.5em}
\section*{References}
\bibliographystyle{iopart-num} 
\bibliography{ref}

\end{document}